# $C_{60}$-BASED COMPOSITES IN VIEW OF TOPOCHEMICAL REACTIONS. I. $C_{60}$ DIMERS AND OLIGOMERS

E.F.Sheka and L.Kh.Shaymardanova

Peoples' Friendship University of Russia
sheka@icp.ac.ru

**ABSTRACT**: The current paper opens a series of papers that are aimed at the determination of barriers that govern the covalent coupling between partners of $C_{60}$-based composites consisting of two or more fullerenes $C_{60}$ ($C_{60}$ dimer and oligomers) (Part 1), $C_{60}$ and single-walled carbon nanotube ($[C_{60}+(4,4)]$ carbon nanobud) (Part 2), and $C_{60}$ and graphene ($[C_{60}+(5,5)]$ and $[C_{60}+(9,8)]$ graphene nanobuds) (Part 3). $C_{60}$ dimers and oligomers are considered in the current paper. The formation of composites is considered from the basic points related to the regioselective chemical reactivity of the fullerene molecule atoms. The dissonance between the predicted trimer and tetramer structures and experimental observations is suggested to evidence the topological nature of the $C_{60}$ oligomerization. The barrier that governs the oligomer formation is determined in terms of the coupling energy $E_{cpl}^{tot}$ and is expanded over two contributions that present the total energy of deformation of the composites' components $E_{def}^{tot}$ and the energy of covalent coupling $E_{cov}^{tot}$. The computations were performed by using the AM1 semiempirical version of unrestricted broken symmetry Hartree-Fock approach.

## 1. Introduction

Recently started manufacturing of nanocarbon-based composite materials pursues well-defined goals to provide the best conditions for the exhibition and practical utilization of extraordinary thermal, mechanic, electronic, and chemical properties of the nanocarbons. Obvious success in achieving the goal in the case of fullerenes, carbon nanotubes (CNTs), and graphene dissolved in different polymers points to great perspectives of a new class of composites and their use in a variety of applications (see Ref.1 and references therein). Since low-concentration solutions of individual fullerenes, CNTs, and graphene flakes can be obtained, one can put a question what can we expect when both fullerene and CNTs or graphene are dissolved simultaneously? First experimental attempts to reach the goal have been successful. A few techniques have been suggested to obtain $C_{60}$-CNT composites in which fullerene is located either inside (see review [2] and references therein) or outside [3-6] the CNT wall. Terms 'nanobud' [4] and peapod [2] were suggested to distinguish the two configurations. However, until recently there has been no information concerning the creation of $C_{60}$-graphene composites while those related to CNT-graphene ones seem to become quite known [7-9].

From the basic standpoint the problem concerning the composite formation addresses the intermolecular interaction (IMI) between the components. In all the cases, the IMI is greatly contributed with the donor-acceptor (DA) interaction since all the above $sp^2$ nanocarbons are simultaneously good donors and acceptors of electron [10-12]. Within the framework of general characteristics of the DA interaction, the IMI term configuration in the ground state depends on the difference of the asymptotes of $E_{int}(A^0B^0)$ and $E_{int}(A^+B^-)$ terms that describe interaction between neutral molecule and molecular ions, $E_{gap} = I_A - \varepsilon_B$. Here $I_A$ and $\varepsilon_B$ present ionization potential and electron affinity of components $A$ and $B$, respectively.

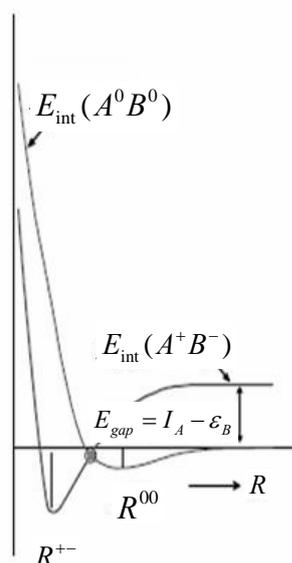

**Figure 1**. Scheme of terms corresponding to the IMI potential of type 1. $(A^0B^0)$ and $(A^+B^-)$ match branches of the terms related to the IMI between neutral molecules and their ions, respectively.

When $E_{gap}$ is as big as it in the case of $C_{60}$ dimers, the IMI term of the ground state has a typical two-well shape shown in Fig.1 [11]. The formation of a stationary product $AB$ at point $R^{(+-)}$ is accompanied by the creation of "intermolecular" chemical bonds between $A$ and $B$ partners. Oppositely, widely spaced neutral moieties form a charge transfer complex $A+B$ in the vicinity of point $R^{(00)}$. In spite of the formation of $AB$ product is energetically profitable, the yield of the relevant reaction when starting from $A+B$ mixture is determined by a barrier that separates $A+B$ and $AB$ products. Since both ionization potentials and electron affinities of $C_{60}$, CNT, and graphene are quite similar by value, the above composites are characterized by practically the same $E_{gap}$. Beside $E_{gap}$, the discussed composites are characterized by the same atom configuration of the contact place. The two factors made it possible to expect a similar behavior of the addition reactions between the composites partner followed by similar profiles of the barrier energy. As turned out, this expectation was not proven by calculations in the current study, thus highlighting that not only $E_{gap}$ but other factors influence the barrier profiles, which makes practical reactions for the three types of composites quite different. To highlight possible reasons that lay behind the finding we have performed an extended computational study aimed at the barrier determination in the case of three types of composites, namely, $C_{60}+C_{60}$, $C_{60}+CNT$, and $C_{60}+$graphene. The computations were performed by using the AM1 semiempirical version of unrestricted broken symmetry Hartree-Fock approach. Since fullerene $C_{60}$ is one of the two members of all mentioned composites, it is reasonable to start their consideration from the simplest and most studied case related to $C_{60}$ dimers and oligomers. Part 2 is devoted to $C_{60}+CNT$ composites while $C_{60}+$graphene ones are considered in Part 3.

## 2. Ground-state term of the $C_{60}+C_{60}$ dyad

Obviously, a successive formation of $C_{60}$ oligomers is governed by the pair $C_{60}$-$C_{60}$

interaction so that energetic parameters of the $C_{60}+C_{60}$ dyad become decisive. According to the scheme in Fig.1, the reaction of the $C_{60}$ dimer formation can be considered as moving the two molecules towards each other, once spaced initially at large intermolecular distance $R$, then equilibrated and coupled as $A+B$ complex in the $R^{00}$ minimum and afterwards achieved minimum at $R^{+-}$ to form tightly bound adduct $AB$. The last stage implies overcoming a barrier, which is followed by the transition from $(A^0 B^0)$ to $(A^+ B^-)$ branch of terms after which Coulomb interaction between molecular ions completes the formation of the final $AB$ adduct at the $R^{+-}$ minimum. As shown [10], neither ionization nor positive charging of $C_{60}$ causes lengthening of the molecule valence bonds more than by 0.02 Å. Therefore, the ion formation is not accompanied by a noticeable shift of the atom equilibrium positions along any internal coordinate and does not cause any significant vibrational excitation during the relevant transition that could have caused the molecule decomposition under ionization. That is why the fullerene dimerization and/or oligomerization occur as a direct addition reaction between non-decomposed molecules [10, 11].

Not only equilibrium configurations $A+B$ and $AB$, which were considered in details earlier (see Ref.11 and references therein), but a continuous transition from one state to the other is the main goal of computations. Following this way we are able to get the barrier profile of the reaction under consideration. Computationally, it seems quite identical to study the profile by either shifting monomer molecules of the $A+B$ complex towards each other, thus contracting the correspondent intermolecular C-C distances, or separating the molecules of the $AB$ product by elongation of the relevant intermolecular C-C bonds. In contrast to the equilibrium configuration $A+B$, which does not critically depend on mutual orientation of the partners, regioselective chemical activity of the $C_{60}$ atoms [12-14] favors a particular combination of the partners' atoms that are involved in the contact zone of the $AB$ product. Since that we shall start from the equilibrium $AB$ structure and shall regularly elongate the relevant intermolecular bonds. The same approach will be used later when considering other composites in this study.

Concerning covalent bonding that involves fullerene $C_{60}$, we should usually proceed from the molecule partial radicalization due to exceeding C-C bonds length a critical value of 1.395Å under which the odd electrons are fully covalently bound forming classical π electron pairs and over which those become effectively unpaired [12]. Thus appeared effectively unpaired electrons form a pool of molecular chemical susceptibility determined by the total number of the unpaired electrons $N_D$. Distributed over the $C_{60}$ molecule atoms by partial number $N_{DA}$, the electron highlights the map of chemical activity of $C_{60}$ in terms of atomic chemical susceptibility (ACS) $N_{DA}$ [12-14]. The variety of the ACS distribution over atoms of the $C_{60}$ molecule is shown in Fig. 2a by different colors that distinguish atoms with different ACS. Among the latter, the most active atoms are shown by light gray. Those are the first targets involved into initial stages of any addition reaction. Following this indication, the initial composition of a pair of the $C_{60}$ molecules shown in Fig.2*a* becomes quite evident. The starting configuration corresponds to $R_{CC}^{st}$ equal to 1.7Å that corresponds to distances between 1-1` and 2-2` target atoms. A bound dumbbell-like dimer is formed (Fig. 2*b*) after the structure optimization aimed at the total energy minimization is completed. It turns out that two monomers within the dimer are contacted via a typical [2+2] cycloaddition of '66' bonds that form a cyclobutane ring [11]. Main electronic characteristics of the $(C_{60})_2$ adduct are presented in Table 1. A detail comparison with other computational data is given elsewhere [11]. A large negative $E_{cpl}^{tot}$ value undoubtedly evidences that $(C_{60})_2$ dimer is actually a typical $AB$ adduct attributed to the $R^{+-}$ minimum on the IMI ground state term in Fig.1.

When $R_{CC}^{st} = 3.07$ Å, optimization of the initial structure leads to a weakly bound pair of molecules spaced by $R_{CC}^{fin} = 4.48$ Å. Monomer molecules preserve their initial configurations and, as seen from Table 1, form a classical charge transfer complex. The fragment composition of the HOMO and LUMO is cross-partitioned; the former should be attributed to Mol 2 while the latter – to Mol 1 just showing that a charge transfer from Mol 2 to Mol 1 occurs when the complex is photoexcited. To obtain the barrier profile, two intermolecular C-C distances, namely, 1-1' and 2-2' separations of the [2+2] cycloaddition in Fig.2b, were stepwise elongated with a constant increment of 0.05Å during the first stage of elongation from 1.57 to 2.22Å and then of 0.1Å during the second stage.

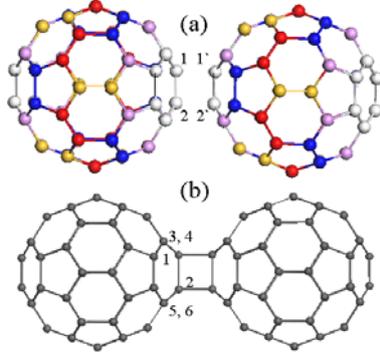

**Figure 2**. (*a*) Start composition of the $C_{60} + C_{60}$ composite. (*b*) Equilibrium structure of the $(C_{60})_2$ dimer; $R_{CC}^{st} = 1.7$ Å, $R_{CC}^{fin} = 1.55$ Å. All the distances correspond to the spacing between 1-1` and 2-2` atoms.

**Table 1**. Electronic characteristics of the $C_{60}+C_{60}$ composites [11]

| Computed quantities<br>singlet, UBS HF AM1 singlet state | Monomer<br>$C_{60}$ | $R_{CC}^{st}$ 1.71 Å | $R_{CC}^{st}$ 3.07 Å |
|---|---|---|---|
| Heat of formation[1], $\Delta H$, kcal/mol | 955.56 | 1868.49 | 1910.60 |
| Coupling energy[2], $E_{cpl}$, kcal/mol | - | -42.63 | -0.52 |
| Ionization potential[3], $I$, eV | 9.86 (8.74[a]) | 9.87 | 9.87 |
| Electron affinity[3], $\varepsilon$, eV | 2.66 (2.69[b]) | 2.62 | 2.66 |
| Dipole moment, $D b$ | 0.01 | 0.001 | 0.001 |
| Squared spin, $(S**2)$[4] | 4.92 | 10.96 | 9.87 |
| Total number of effectively non-paired electrons, $N_D$ | 9.84 | 21.93 | 19.75 |
| Gained charge to *Mol 1* | - | 0.0 | 0.0 |
| Transferred charge from *Mol 2* | - | 0.0 | 0.0 |
| Symmetry | $C_i$ | $C_{2h}$ | $C_i$ |
| HOMO, fragment compositions, $\eta$ | - | $\eta_{Mol1}$ =61.8%<br>$\eta_{Mol2}$=38.1% | $\eta_{Mol1}$ =0%<br>$\eta_{Mol2}$=100% |
| LUMO, fragment compositions, $\eta$ | - | $\eta_{Mol1}$ =83.9%<br>$\eta_{Mol2}$=15.8% | $\eta_{Mol1}$ =100%<br>$\eta_{Mol2}$=0% |

Note 1: Molecular energies are by heats of formation $\Delta H$ determined as $\Delta H = E_{tot} - \sum_A (E_{elec}^A + EHEAT^A)$. Here $E_{tot} = E_{elec} + E_{nuc}$, while $E_{elec}$ and $E_{nuc}$ are the electron and core energies. $E_{elec}^A$ and $EHEAT^A$ are electron energy and heat of formation of an isolated atom, respectively.
Note 2: Coupling energy is determined following Eq.(1).
Note 3: Here $I$ and $\varepsilon$ correspond to the energies of HOMO and LUMO, respectively, just inverted by sign. Experimental data for the relevant orbitals are taken from [15] (*a*) and [16] (*b*).
Note 4: Non-zero squared spin evidences a spin-mixed of the molecule's singlet state [12-14].

Figure 3 presents the barrier profile of the $C_{60}$ dimerization in terms of the total coupling energy $E_{cpl}^{tot}$ determined as

$$E_{cpl}^{tot}(R_{CC}) = \Delta H_{dim}(R_{CC}) - 2\Delta H_{mon}^{eq} \qquad (1)$$

where $\Delta H_{dim}(R_{CC})$ and $\Delta H_{mon}^{eq}$ present heats of formation of dimer at the current intermolecular distance $R_{CC}$ and of monomer in equilibrium, respectively. This energy is evidently complex by nature since at least two components contribute into its value, namely the energy of both monomers deformation $E_{def}^{tot}$ and the energy of the covalent coupling $E_{cov}^{tot}$ between the monomers. The former component related to both monomer molecules can be determined as

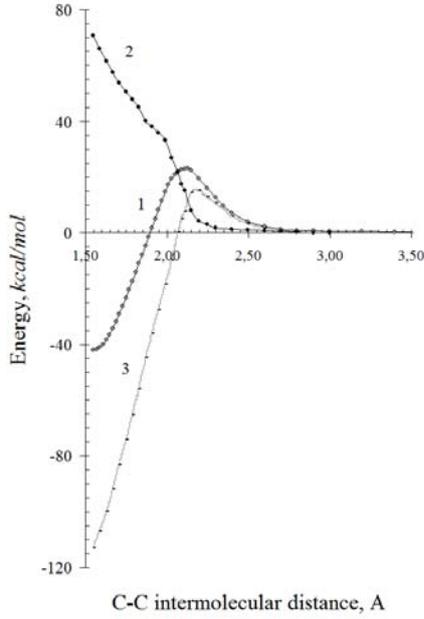

$$E_{def}^{tot}(R_{CC}) = \Delta H_{mon1}(R_{CC}) + \Delta H_{mon2}(R_{CC}) - 2\Delta H_{mon}^{eq}. \qquad (2)$$

**Figure 3.** Profile of the barrier of the $(C_{60})_2$ dimer decomposition. 1. $E_{cpl}^{tot}$; 2. $E_{def}^{tot}$; 3. $E_{cov}^{tot}$.

Here $\Delta H_{mon1}(R_{CC})$ and $\Delta H_{mon2}(R_{CC})$ present heats of formations of both one-point-geometry monomer molecules in the structure that correspond to the $C_{60}+C_{60}$ dyad at $R_{CC}$ intermolecular distance. The second component $E_{cov}^{tot}$ we determine as

$$E_{cov}^{tot}(R_{CC}) = E_{cpl}^{tot}(R_{CC}) - E_{def}^{tot}(R_{CC}). \qquad (3)$$

The discussed distance-dependent energies are shown in Fig.3. As seen in the figure, the deformation energy is the largest in equilibrium dimer and then steadily decreases when $R_{CC}$ grows, but being positive, until approaching zero when monomers are spaced by more than 3Å. Similarly, the energy of the covalent coupling is the largest in equilibrium dimer then steadily decreasing by absolute value being negative and showing a clearly vivid maximum that coincides with that of $E_{cpl}^{tot}(R_{CC})$ after which it falls by absolute value, changes sign in the vicinity of $R^{00}$ ~4.4Å, once being small by absolute value, and then approaches zero for largely spaced monomers. Referring to schemes of electronic terms in Fig.1, one should accept that this is the energy $E_{cov}^{tot}$ that should be attributed to the *netto* barrier profile. However, the energy $E_{cpl}^{tot}(R_{CC})$ as a *brutto* barrier profile will obviously govern the decomposition of fullerene dimers in practice.

## 3. $C_{60}$ oligomers and topochemical reactions

The detail study of the IMI term's profiles, which governs $C_{60}$ dimerization discussed above, opens way to throw light on peculiarities of the $C_{60}$ oligomerization. Reasonably, the oligomerization can be computationally considered as a stepwise reaction $(C_{60})_n = (C_{60})_{n-1} + C_{60}$ for which the IMI term of the $[(C_{60})_{n-1} + C_{60}]$ dyad controls the formation of the final product $(C_{60})_n$. Actually, one cannot exclude more complex scheme, such as $(C_{60})_n = (C_{60})_m + (C_{60})_k$ where $m+k=n$, $(C_{60})_n = (C_{60})_m + (C_{60})_k + (C_{60})_l$ where $m+k+l=n$, and so forth. However, all the schemes are subordinated to common regularities whose main peculiarities can be considered for the $(C_{60})_{n-1} + C_{60}$ dyad as an example.

If accept that the type of IMI terms is mainly determined by $E_{gap} = I_A - \varepsilon_B$ as discussed in Section 1, passing to oligomers one faces a peculiar situation characteristic for fullerenes. The matter is that both ionization potential and electronic affinity of $(C_{60})_n$ oligomer only slightly depend on $n$ and practically coincide with those related to monomer molecule. This can be seen in Table 1 for dimer and has been computationally justified for oligomers of complex structure characterized by $n$ varying up to 10 [17]. Consequently, the IMI term of the $C_{60} + C_{60}$ dyad determines the behavior of both $(C_{60})_{n-1} + C_{60}$ and $(C_{60})_m + (C_{60})_k$ dyads at each successive step of oligomerization. As in the case of dimers, two products, namely either $(C_{60})_{n-1} + C_{60}$ or $(C_{60})_m + (C_{60})_k$ charge transfer complexes and $(C_{60})_n$ oligomer will correspond to equilibrium positions at $R^{00}$ and $R^{+-}$ minima of the IMI term. Following this suggestion and taking into account concepts of computational chemistry of fullerenes, one can suggest a definite scheme of the expected successive oligomerization of $C_{60}$ molecules when going, say, from dimer to tetramer within the $(C_{60})_n = (C_{60})_{n-1} + C_{60}$ oligomerization scheme, as shown in Fig.4.

According to the ACS map of dimer $(C_{60})_2$, there are four pairs of high-rank-$N_{DA}$ atoms that are marked by red balls in the right low corner of Fig.4. The first two pairs combine the most reactive atoms adjacent to the cycloaddition (below, contact-adjacent or *ca* atoms) (see atoms 3, 4 and 5, 6 in Fig. 2). Next by reactivity four atoms are located in equatorial planes of both monomers (below, equatorial or *eq* atoms). In spite of high chemical reactivity of *ca* atoms, those are not accessible in due course of the further oligomerization so that *eq* atoms of both monomers are actual targets. Following these ACS indication, a right-angle-triangle trimer ($90^0$-trimer) must be produced. Therefore, not the 'pearl necklace' configuration, intuitively suggested as the most expected for $C_{60}$ oligomerization [18], but more complicated 2D one is favorable for trimerization. Moreover, edge atoms located at both ends of the dimer horizontal axis are characterized by the least values on the ACS map, so that the formation of a linear trimer is evidently less probable.

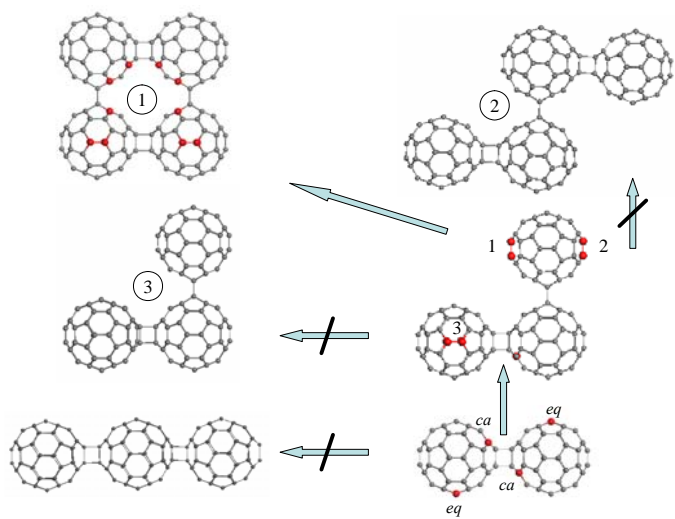

**Figure 4**. Stepwise oligomerization of $C_{60}$ from dimer to tetramer. Equilibrium structures. Crossed arrows indicate unfavorable continuations. Coupling energies constitute –42.23 kcal/mol (dimer); -74.73 kcal/mol (trimer); -164.63 kcal/mol (tetramer 1); -13.84 kcal/mol (tetramer 2); -117.66 kcal/mol (tetramer 3).

Similarly, the high-rank $N_{DA}$ atoms of trimer as seen in Fig.4 form an incomplete *ca* pair of the highest activity and three pairs of *eq* atoms of comparable activity. Three tetramer compositions that follow from the high-rank ACS indication related to trimer are shown in Fig.4. None of them belongs to the 'pearl necklace' family thus presenting 2D tetramers 1 and 3 and 3D tetramer 3. Among the latter, tetramer 1 possesses the least energy and is expected to continue the oligomerization offering its high-rank $N_{DA}$ atoms, marked by red balls, as targets for the next $C_{60}$ addition. Those form 6 pairs of the most active *ca* atoms and four pairs of *eq* atoms, position of which dictates the continuation of oligomerization as the formation of 3D configurations of pentamers.

Before passing to comparison with experimental data, it should be noted that the considered way of the oligomerization is related to the addition reaction that occurred between partners in vacuum without any constrains on their orientation. The latter may be adapted if necessary by free motion of the partners. As turned out, experimental situation significantly differs from the considered. Performed experiments differ quite drastically relating to the study of $C_{60}$ clusters in the gas phase (see [19, 20] and references therein), solutions [21], solid films and powders [22-26], and solids [27-29]. All the studies clearly evidence the formation of $(C_{60})_n$ clusters with $n$=2, 3, 4, and more while the information concerning the cluster's structure is rather scarce. The best situation concerns dimers whose dumbbell-like structure was proven by many ways (see [11] and references therein). Applying to trimers, there are two sources of information related to drastically different ways of their production that conditionally can be denoted as 'chemical' [23] and 'physical' [24, 25]. 'Chemical' experiments [23] dealt with $C_{180}$ species produces in due course of solid state mechanochemical reaction of $C_{60}$ with 4-aminopyridine under high-speed vibration milling conditions. The final product, investigated by using STM, has exhibited two fractions (A and B in ratio of ~ 5:4), the former predominantly (~60%) consisting of $90^0$-trimers while 100% of the latter is presented by cyclic $60^0$- trimers. 'Physical' experiments deal with trimers produced under photoilumination of either $C_{60}$ films preliminary deposited on some substrates [25-27] or pristine $C_{60}$ crystal [29]. Linear three-ball chains were observed in these studies only.

Evidently, none of the two experimental procedures is similar to the computationally considered. From this viewpoint, even a predominant presence of $90^0$-trimers within fraction A of 'chemical' experiment cannot be considered as a doubtless proof of compositions predicted by the ACS-guided covalent chemistry of $C_{60}$ since three other $108^0$-, $120^0$-, and $144^0$-trimers, observed within fraction A, as well as $60^0$- trimers of fraction B [23] seriously contradict the scheme shown in Fig.4. Even more stronger contradiction is revealed by comparison with 'physical' experiments that exhibit only linear 'pearl necklace' trimers, the less probable in accordance with the ACS-based analysis. At the same time, compositionally simple 'physical' findings evidently connect the trimerization events with a predetermined molecular packing. As turns out, the observation of similar events is not rare and those are usually related to *topochemical reactions* whose occurring is controlled by the reactants packing. Thus, firstly observed for $C_{70}$ fullerene, the linear trimerization as well as linear polymerization of the molecules in solid state was explained [30] as topochemical reaction controlled by monomer crystal packing when "the alignment of molecules and their presumed orientational mobility facilitate polymerization via spatial adjustment of reactive double bonds of neighboring cages". Therefore, all the observed 'physical' polymerization events that resulted in the production of linear orthorhombic crystalline modification of polymerized $C_{60}$ [30] should be obviously attributed to the relevant topochemical reactions. On the other hand, the mechanochemical reaction, responsible for triangle trimers, evidently has a topological odor as well thus providing formation of differently configured trimers due to obvious anisotropy of the stress application to the pristine $C_{60}$ crystal under milling.

The situation with tetramers is not simple as well. The first suggestion concerning a close structure of $(C_{60})_4$ clusters analogous to tetramer 1 in Fig.4 was issued for clusters in solutions [21]. Later on the suggestion was supported by the analysis of Raman spectra of photoiluminated $C_{60}$ powders [22] as well as direct STM observation of deposited $(C_{60})_n$ clusters on the (111) surface of gold [28]. In spite of a seemingly favorable fitting of the experimental data to predicted ones, one has to take into account that both experiments are performed under evident conditions that favor topochemical reactions. Thus, not only $(C_{60})_4$ clusters deposited on the gold surface have 2D tetragonal shape similar to tetramer 1 but all other $(C_{60})_n$ clusters with *n*>4 clearly exhibit close 2D configurations in contrast to the predicted 3D ones for pentamers and higher oligomers, which follows from the high-rank $N_{DA}$ ACS indication concerning the tetramer 1 structure shown in Fig.4. The tendency of $(C_{60})_n$ clusters to be inclined in topochemical reaction is obviously realized in 1D orthorhombic as well as 2D tetragonal and rhombohedral configurations of polymerized $C_{60}$ crystals, whose production is controlled by varying one-direction contraction of the pristine $C_{60}$ crystal structure at high pressures and temperatures [29].

## 4. Conclusive remarks about the character of chemical reactions typical to fullerene

Fullerenes are exclusive molecular species that cannot be assigned to a particular class of compounds. Consequently, not a certain class of chemical reactions, as traditionally is taking place, but a large set of transformations of different types are characteristic for them. Two main reasons lay the foundation of the behavior: high chemical susceptibility caused by effectively unpaired electrons and high donor and acceptor abilities of the species. Taking together theses factors enlarge considerably the variety of reactions to be occurred with fullerene participation from traditional addition reactions to complicated ones governed by the DA interaction mainly. It seems quite reasonable to suggest the topochemical reactions among the latter.

Traditional addition reaction is not a proper term related to fullerene-based reactions since kinetic of the latter is rigidly controlled by the regioselectivity caused by the effectively unpaired electron distribution over atoms. Seemingly, a multi-fold isomerism of the fullerene molecules is needed to be taken into account as well. On the other hand, DA reactions cannot be assigned as traditional as well since their particular features are connected with a multi-well structure of the ground state energy term of the relevant binary system. A large variety of the DA reactions of fullerenes can be explained by changing the mutual disposition of the well minima. Another feature of the fullerene-involving DA reactions concerns the formation of not only hetero-reactant but also homo-reactant products. The latter implies dimers and higher fullerene oligomers. While in the hetero-reactant case fullerenes are electron acceptors, the species play both acceptor and donor role in the homo-reactant case.

This bi-mode DA ability is characteristic for not only fullerenes but for other $sp^2$ nanocarbons such as CNT and graphene. Consequently, the reactions, or better to say, IMI between components of dyads such as fullerene+CNT, fullerene+graphene, CNT+CNT, CNT+graphene as well as more complicated triads, tetrads and so forth should be considered in terms of DA peculiarities discussed above.

### References

1. *Sheka, E. F., Chernozatonskii, L. A.* Graphene-carbon nanotubes composites //J. Compt. Theor. Nanosci., 2010, **7**, 1814.

2. *Giacalone, F., Martín, N.* New Concepts and Applications in the Macromolecular Chemistry of Fullerenes // Adv. Mat. 2010, **22**, 4220.
3. *Li, X., Liu, L., Qin, Y., Wu, W., Guo, Zh.-X., Dai, L., and Zhu, D.* $C_{60}$ modified single-walled carbon nanotubes // Chem. Phys. Lett. 2003, **377**, 32.
4. *Nasibulin, A. G., Anisimov, A. S., Pikhitsa, P. V., Jiang, H., Brown, D. P., Choi, M., and Kauppinen, E. I.* Investigations of NanoBud formation // Chem. Phy. Lett. 2007, **446**, 109.
5. *Nasibulin, A. G., Pikhitsa, P. V., Jiang, H., Brown, D. P., Krasheninnikov, A.V., Anisimov, A. S., Queipo, P., Moisala, A., Gonzalez, D., Lientschnig, G., Hassanien, A., Shandakov, S.D., Lolli, G., Resasco, D.E., Choi, M., Tománek, D., and Kauppinen, E. I.* A novel hybrid carbon material // Nature Nanotechn. 2007, **2**, 156.
6. *Li, C., Chen, Y., Wang, Y., Iqbal, Z., Chhowallab, M. and Mitra, S.* A fullerene–single wall carbon nanotube complex for polymer bulk heterojunction photovoltaic cells // J. Mat. Chem. 2007, **17**, 2406.
7. *Kondo, D., Sato, S., and Awano, Y.*. Self-organization of novel carbon composite structure: Graphene multi-layers combined perpendicularly with aligned carbon nanotubes // Appl. Phys. Express 2008, **1**, 074003.
8. *Lee, E. J. H., Zhi, L., Burghard, M., Müllen, K., Kern, K.* Electrical Properties and Photoconductivity of Stacked-Graphene Carbon Nanotubes //Adv. Mat. 2010, **22**, 1854.
9. *Jousseaume,V.; Cuzzocrea, J.; Bernier, N.; Renard V. T.* Few graphene layers/carbon nanotube composites grown at complementary-metal-oxide-semiconductor compatible temperature //Appl. Phys. Lett. **2011**, *98*, 123103.
10. *Sheka, E.F.* Intermolecular interaction in $C_{60}$-based donor acceptor complexes //Int. J. Quant. Chem. **2004**, *100*, 388-406.
11. *Sheka, E.F.* Donor-acceptor interaction and fullerene $C_{60}$ dimerization //Chem. Phys. Lett. **2007**, *438*, 119-126.
12. *Sheka, E.F.* Fullerene Nanoscience : Nanochemistry, Nanomedicine, Nanophotonics, Nanomagnetism (Boca Raton: Taylor and Francis) **2011**.
13. *Sheka, E.F., Zayetz, V.A.* The radical nature of fullerene and its chemical activity //Russ Journ. Phys. Chem. **2005**, *79,* 2009-2014.
14. *Sheka, E.F.* Chemical susceptibility of fullerenes in view of Hartree-Fock approach //Int. Journ. Quant. Chem. **2007**, *107***,** 2803-2816.
15. *Weaver, J.H., Martins, J.L., Komeda, T., Chen. Y., Ohno, T.R., Kroll, G.H., Troullier, N., Haufler, R., and Smalley, R.E.* Electronic structure of solid $C_{60}$: Experiment and theory // Phys. Rev. Lett. 1991, **66**, 1741.
16. *Wang, X.-B., Ding, C.-F., and Wang, L.-S.* High resolution photoelectron spectroscopy of $C_{60}^{-}$. //J. Chem. Phys. 1999, **110**, 8217.
17. *Sheka, E.F., Zaets, V.A., and Ginzburg, I.Ya*. Nanostructural magnetism of polymeric fullerene crystals.//J. Exp. Theor. Phys. 2006,**103**, 728.
18. *Fischer, J.T.* Fullerene polymers from solid precursors? //Science 1994, **264**, 1548.
19. *Hedén, M., Hansen, K., and Campbell, E.E.B*. Molecular fusion of $(C_{60})_N$ clusters in the gas phase after femtosecond laser irradiation //Phys. Rev. 2005, **71A**, 055201.
20. *Enders, A., Malinowski, N., Ievlev, D., Zurek, E*. Magic alkali-fullerene compound clusters of extreme thermal stability //J. Chem. Phys. 2006, **125**, 191102.
21. *Sun, Y.-P., Ma, B., Bunker, C. E., and Liu, B.* All-carbon polymers (polyfullerenes) from photochemical reactions of fullerene clusters in room-temperature solvent mixtures //J. Am. Chem. Soc. 1995, **117**, 12705.
22. *T. Pusztai, T., Oszla´nyi, G., Faigel, G., Kamaras, K., Granasy, L., and Pekker, S.* Bulk structure of phototransformed $C_{60}$ //Solid St. Commun. 1999, **111**, 595.

23. *Kunitake, M., Uemura, S., Ito, O., Fujiwara, K., Murata, Y. and Komatsu, K.* Structural analysis of $C_{60}$ trimers by direct observation with scanning tunneling microscopy //Angew. Chem. Int. Ed. 2002, **41**, 969.
24. *Nakaya, M., Kuwahara, Y., Aono, M. and Nakayama, T.* Reversibility-controlled single molecular level chemical reaction in a $C_{60}$ monolayer via ionization induced by scanning transmission microscopy //Small. 2008, **4**, 538.
25. *Nakayama, T., Onoe, J., Nakatsuji, K., Nakamura, J., Takeuchi, T., and Aono, M.* Photoinduced products in a $C_{60}$ monolayer on Si(111)($\sqrt{3}x\sqrt{3}$)-Ag: An STM study //Surf. Rev. Lett. 1999, **6**, 1073.
26. *Ecklund, P.C., Rao, A.M., Zhou, P., Wang, Y., and Holden, J.M.* Photochemical transformation of $C_{60}$ and $C_{70}$ films //Thin Solid Films 1995, **257**, 185.
27. *Wang, G., Komatsu, K., Murata, Y., and Shiro, M.* Synthesis and X-ray structure of dumb-bell-shaped $C_{120}$ //Nature, 1997, **387**, 583.
28. *Zhang, X., Tang, L., and Guo, Q.* Low-temperature growth of $C_{60}$ monolayers on Au(111): Island orientation control with site-selective nucleation //J. Phys. Chem., 2010, **114C**, 6433.
29. *Núñez-Regueiro, M., Markes, L., Hodeau, J.-L., Béthoux, O., and Perroux, M.* Polymerized fullerite structures //Phys. Rev. Lett., 1995, **74**, 278.
30. *Soldatov, A.V., Roth, G., Dzyabchenko, A., Johnels, D., Lebedkin, S., Meingast,C., Sundqvist, B., Haluska, M., and Kuzmany, H.* Topochemical polymerization of $C_{70}$ controlled by monomer crystal packing //Science, 2001, **293**, 680.